\documentclass[twoside]{article}
\usepackage{fleqn,espcrc2}

\title{Supersymmetry on Lattice Using Ginsparg-Wilson Relation}

\author{Hiroto. So\address[Niigata]{%
     Department of Physics, Niigata University,
     Ikarashi 2-8050, Niigata, 950-2181, Japan},
  Naoya Ukita\addressmark[Niigata]\thanks{presented by N. Ukita 
     at Lattice 2000, Bangalore, India.}}

\begin{document}

\begin{abstract}
 The Ginsparg-Wilson(G-W) relation for  chiral symmetry is extended for  
 a supersymmetrical(SUSY) case on a lattice.  
 It is possible to define exact lattice supersymmetry which are
 devided into two different cases according to using difference operators.
 $U(1)_R$ symmetry on the lattice is also realized as one of exact
 symmetries. For an application, the extended G-W relation is given
 for a two-dimensional model with chiral multiplets.
\end{abstract}

\maketitle

\setcounter{footnote}{0}

\section{Introduction}

 Although  numerical QCD studies have succeeded  in computer simulations 
 by  the lattice framework,  the  problem of 
 chiral fermion\cite{nogo}  is remained. 
 Kaplan's domain wall fermion\cite{kaplan}  and 
 the overlap formula\cite{overlap}   as the hamiltonian formalism is 
 a nice idea if  the chiral charge  is  
 sufficiently separate because the total theory is vector-like and consistent 
 with  the no-go theorem\cite{nogo}.   
 Recently, Neuberger\cite{neu}  found that the Dirac operator by his overlap 
 formula is satisfied with the Ginsparg-Wilson(G-W) 
 relation\cite{gw}.  Furthermore, although the Dirac operator 
 is not chiral-invariant,  the new symmetry found by 
 L{\"u}scher \cite{lu} is a modified version  of  chiral symmetry.  
 It is noted that the usual path-integral measure  is not invariant 
  under the new symmetry  and the  chiral anomaly is generated.

 Lattice formulations of supersymmetry(SUSY)
 were attempted by several authors\cite{ss} - \cite{bie}.  
 But there are many problems such as definition of SUSY on lattice, 
 non-locality and breakdown of Leibniz rule, 
 because the SUSY tansformation  need  not only  chiral fermions 
 but also derivative operators. 
  In this article, we explain  a SUSY-extended  G-W relation\cite{su}.
A  new symmetry for lattice SUSY can be defined without 
 any ambiguities in replacing   derivative  operators with difference
 ones.  As the compensation, it  is necessary
 for  the block-spin funcion constructing  a lattice theory.   
  As a example, a SUSY free action  in two-dimension 
 is constructed.
   In addition to  fermionic and bosonic G-W relations used by
 Aoyama-Kikukawa\cite{ak},  we can derive the further two relations
 among fermions and bosons for the lattice action.   

\section{Derivation of SUSY Ginsparg-Wilson relation}

 The original G-W relation was derived as an identity 
 for  the Gaussian type  effective action.
  It should be recognized as a kind of Ward-Takahashi identity
 for the chiral transformation of the Gaussian type 
 effective action  and as the property of the lattice theory near the continuum  limit with a chiral symmetry\cite{ksu}.

  Due to the no-go theorem\cite{nogo}, we must prepare two  chiral 
 matter fields.
  A SUSY transformations for  two chiral-multiplets, $\Phi_j = 
 (\phi_j, \psi_j, F_j)^{\rm T}$ $j=1,2$   in the continuumm theory is 
 defined as

\begin{equation}
\delta_{\epsilon} \Phi_j =  Q(\epsilon , \bar{\epsilon})\Phi_j 
\end{equation}

\begin{equation}
\delta_{\epsilon} \bar{\Phi}_j = \bar{\Phi}_j  \bar{Q}(\epsilon , \bar{\epsilon})
\end{equation}

\noindent 
where an  Euclidean space is considered because we construct 
 the Euclidean lattice theory. 
 The dependency on index $j$ is trivial and we suppress it from now.
We begin with a continuum theory, and define its regularized theory on a 
cubic lattice by performing a block-spin transformation. 
 A lattice point is expressed by an integer vector, 
\{$n_{\mu}a$\}, where $a$ is lattice constant. We take for simplicity $a = 1$. 
 The block-spin transformation from $\Phi (x)$ to $\Phi_n$ is given by

\begin{equation}
\Phi_n  \sim \displaystyle{ \int} {\rm d}x f_n(x) \Phi (x) ~~ \equiv 
~~\langle f_n, \Phi \rangle 
\end{equation}

\begin{equation}
\bar{\Phi}_n  \sim \displaystyle{ \int} {\rm d}x f_n(x) \bar{\Phi} (x)
 ~~ \equiv 
~~\langle f_n, \bar{\Phi} \rangle 
\end{equation}

\noindent 
where $f_n(x)$ is a block-spin function with finite support around
 $x_{\mu} = n_{\mu}$, and co-moving with $x_{\mu} - n_{\mu}$: $f_n(x) =
 f(x-n)$.  $\langle , \rangle$ implies the usual inner product in a function 
space.

According to  Ginsparg
and Wilson,\cite{gw}
we may define a Gaussian effective action $A_{\rm eff}$ by using a SUSY
action $A_c$ in the continuum theory\footnote{Since we have two chiral 
multiplets, it is possible to construct a Dirac mass term.}:

$$
\displaystyle{\exp (-A_{\rm eff}[\Psi_n, \bar{\Psi}_n])} 
$$

\begin{equation}
\begin{array}{ll}
  =&  \int {\cal D}\Phi (x)  {\cal D}\bar{\Phi} (x)
\exp ( - A_c [\Phi, \bar{\Phi}]  \\
   & \\
   &- \sum_{n,m}(\bar{\Psi}_n - \bar{\Phi}_n)
\alpha_{n,m}(\Psi_m - \Phi_m)) . 
\end{array}
\end{equation}

\noindent
Here $\alpha_{n,m}$ is a matrix acting on the  multiplet, $\Psi_n$,

\begin{eqnarray}
\alpha_{n,m} = \alpha \  \delta_{n,m} \left(\begin{array}{rrr}
0 & 0 & 1 \\
0 & V & 0 \\
 1 & 0 & 0 
\end{array}
\right)        \label{v}
\end{eqnarray}

\noindent
where $V$ is some anti-symmetric matrix determined by a mass-term
Lagrangian.
 This  $\alpha_{n,m}$ term is a SUSY invariant mass term with a large
mass   $O(a^{-1})$.
 In the na{\"\i}ve sense, our effective  action, $A_{\rm eff}$  is SUSY
 invariant but  the transformation is modified as seen later.

We may define na{\"\i}ve SUSY on the lattice:

\begin{equation}
\begin{array}{lll}
\delta^N_{\epsilon} \Phi_n & = & \displaystyle{\int }  f_n (x) \delta \Phi (x)
{\rm d}x \\
 & & \\
 & = & Q_L(\epsilon , \bar{\epsilon};\overrightarrow{\bigtriangledown})\Phi_n 
\end{array}
\end{equation}

\begin{equation}
\begin{array}{lll}
\delta^N_{\epsilon} \bar{\Phi}_n & = & \displaystyle{\int }  f_n (x) \delta \bar{\Phi} (x)
{\rm d}x \\
 & & \\
 & = &  \bar{\Phi}_n \bar{Q}_L(\epsilon , \bar{\epsilon};
 \overleftarrow{\bigtriangledown})
\end{array}
\end{equation}

\noindent
This is just an interpretation from continuum SUSY to lattice SUSY, where
 a derivative operator in the continuum SUSY 
 is replaced by a difference operator in the lattice SUSY 
using the relation $\partial_{\mu}f_{n}(x)= - 
\overrightarrow{\bigtriangledown}_{\mu}f_{n}(x)$. 
 Although the explicit form of the  difference operator depends on the 
 block-spin function, 
  it is possible to choose  the reasonable function  in 
 the realization of  lattice SUSY. 
Under this na{\"\i}ve lattice SUSY transformation,  
 our effective action changes by

$$
\exp (-A_{\rm eff}[\Psi ', \bar{\Psi} ']) 
$$
$
\begin{array}{cl}
  = & \displaystyle{\int} {\cal D}\Phi (x) {\cal D}\bar{\Phi} (x)
 \exp ( - A_c [\Phi , \bar{\Phi}]   \\
    & \\
    &  - (\bar{\Psi}' - \bar{\Phi})\alpha(\Psi' - \Phi))\\
    & \\
 =  &  \displaystyle{\int} {\cal D}\Phi (x) {\cal D} \bar{\Phi} (x)
       \exp ( - A_c [\Phi ', \bar{\Phi} '] \\
    &\\
    &  - (\bar{\Psi} - \bar{\Phi}') e^{\bar{Q}_L}\alpha
 e^{Q_L}(\Psi - \Phi'))
\end{array}$

\noindent
where the lattice site and the spinor indices are both omitted. It is
assumed that  $A_c$ 
 is invariant under  SUSY transformation in the continuum theory,

\begin{equation}
A_c[\Phi , \bar{\Phi}] = A_c[\Phi ', \bar{\Phi} '] .
\end{equation}

\noindent
Althogh the path-integral measure is  na{\"\i}vely unchanged

\begin{equation}
{\cal D}\Phi'{\cal D}\bar{\Phi}' = {\cal D}\Phi {\cal D}\bar{\Phi} ,
\end{equation}

\noindent
we include the Jacobian factor in the following calculation
 and shall find the relation with the anomaly\cite{fuji}.

Let us derive a SUSY extension of the G-W relation for a free theory
described by 

\begin{equation}
A_{\rm eff}[\Psi , \bar{\Psi}] = \sum_{n,m} 
\bar{\Psi}_n S_{(n,m)} \Psi_m .
\end{equation}

\noindent
Under the naive SUSY, it transforms as

$$
\exp (-A_{\rm eff}[\Psi , \bar{\Psi}])\{1 -  \bar{\Psi} (S Q_L +
\bar{Q}_L S)\Psi \} 
$$

\begin{equation}
\begin{array}{ll}
= & \{1 + \delta J + {\rm str}~(Q_L + \bar{Q}_L) \\
  & \\
  &  - {\rm str}~(  Q_L\alpha^{-1}S + S\alpha^{-1}\bar{Q}_L ) \\
  &  \\
  & -\bar{\Psi}S 
   ( Q_L\alpha^{-1} + \alpha^{-1}\bar{Q}_L ) S\Psi \} \\
  & \\ 
  & \times\exp (-A_{\rm eff}[\Psi , \bar{\Psi}]),  
\end{array}
\end{equation}

\noindent
where 
 $\delta J$ comes from 
 a Jacobian factor.
 So, we can get following two relations;

\begin{equation}
\begin{array}{lll}
 \delta J &= &{\rm str}~( Q_L\alpha^{-1}S +  S\alpha^{-1} \bar{Q}_L ) \\
           & & \\
           & &-{\rm str}~( Q_L + \bar{Q}_L),
\end{array}
\end{equation}

\noindent 
and

\begin{equation}
\begin{array}{l}
 \bar{\Psi} (S Q_L + \bar{Q}_L S)\Psi \\
\\
 \,\,\,\,\,= 
\bar{\Psi}S( Q_L\alpha^{-1} + \alpha^{-1}\bar{Q}_L)S\Psi  . \label{gw2}
\end{array}
\end{equation}

\noindent
These are SUSY extended GW relations. Note that the  right
hand side of Eq.(\ref{gw2}) vanishes if the difference operator
$(\overrightarrow{\bigtriangledown})$ of 
$Q_{L}(\overrightarrow{\bigtriangledown})$, 
$\bar{Q}_{L}(\overleftarrow{\bigtriangledown})$ 
is symmetric. 
Therefore eq.(\ref{gw2}) is divided into two cases according to the
difference operator $(\overrightarrow{\bigtriangledown})$.
For a symmetric difference operator
$\overrightarrow{\bigtriangledown_s}$,
the above relations suggest us to difine

\begin{equation}
q_s \equiv Q_L(\overrightarrow{\bigtriangledown_s})
\end{equation}
\begin{equation}
\bar{q}_s \equiv \bar{Q}_L(\overleftarrow{\bigtriangledown_s})
\end{equation}

\noindent
under which we can show the invariance of our effective action:

\begin{equation}
\delta A_{\rm eff}=\bar\Psi(S q_{s} + \bar{q}_{s} S)\Psi  = 0. 
\end{equation}

\noindent
For a non-symmetric  difference operator
$\overrightarrow{\bigtriangledown}$,

\begin{equation}
q \equiv Q_L(\overrightarrow{\bigtriangledown}) -  
    Q_L(\overrightarrow{\bigtriangledown})\alpha^{-1}S
\end{equation}
\begin{equation}
\bar{q} \equiv \bar{Q}_L(\overleftarrow{\bigtriangledown}) -  
    S\alpha^{-1}\bar{Q}_L(\overleftarrow{\bigtriangledown})
\end{equation}

\noindent
and

\begin{equation}
\delta A_{\rm eff}=\bar\Psi(S q + \bar{q} S)\Psi  = 0. 
\end{equation}

\noindent
This is a SUSY extension of  L{\"u}scher's symmetry\cite{lu}.

 For $U(1)_R$ charge  $Q_R$, we can find the G-W relations 
similar to the SUSY case:

\begin{equation}
\begin{array}{lll}
 \delta J_R &=& {\rm str}~( Q_R\alpha^{-1}S +  S\alpha^{-1} \bar{Q}_R )
 \\
 & & \\
 & &  -{\rm str}~( Q_R + \bar{Q}_R),
\end{array} 
\end{equation}

\noindent
and

\begin{equation}
\begin{array}{l}
 \bar{\Psi} (S Q_R + \bar{Q}_R S)\Psi \\
 \\
\,\,\,\,\, = 
\bar{\Psi}S( Q_R\alpha^{-1} + \alpha^{-1}\bar{Q}_R)S\Psi  .
\end{array}
\end{equation}

\noindent 
The conserved '$U(1)_R$' charge \cite{ak},

\begin{equation}
q_R \equiv Q_R(1 - \alpha^{-1}S),
\end{equation}
 
\noindent
can be also found as one of exact symmetries on the lattice.

\section{An example: 2-Dimensional chiral multiplets}

 We consider two  chiral-multiplets $\Phi_j$, $j=1,2$, consisted of real scalars $\phi_j$,  auxiliary fields $F_j$ and complex Weyl spinors $\chi_j$. 
 There are arranged in
  a complex multiplet $\Phi = (\phi_1 + i\phi_2, F_1 + iF_2;
 \chi_1 + i\chi_2, \chi^*_1 + i \chi^*_2)^{\rm T} \equiv
 (\phi, F;  \chi, \bar{\chi})^{\rm T}$ and its congugate $\bar{\Phi} =
 (\phi_1 -i\phi_2,  F_1 - iF_2; \chi_1 - i\chi_2,
\chi^*_1 - i \chi^*_2) ) \equiv (\phi^*, F^{*}; \bar{\chi}^{\dagger}, 
\chi^{\dagger}) $.
 We define N=1 SUSY transformation;

\begin{equation}
 \left\{ \begin{array}{l}
\delta_\epsilon \phi = i (\epsilon ^* \chi  + \epsilon  \bar{\chi})\\
\delta_\epsilon F =  - 2  \epsilon  \partial_{\bar{z}} \chi + 2  \epsilon ^* 
\partial_z \bar{\chi} \\
\delta_\epsilon \chi = - 2   \epsilon  ^* \partial_z \phi + i \epsilon F \\
\delta_\epsilon \bar{\chi} =  - 2   \epsilon  \partial_{\bar{z}} \phi  -
 i\epsilon ^* F .
\end{array} \right. 
\end{equation}

\noindent
We consider a SUSY-invariant massless Lagrangian:

\begin{equation}
{\cal L} =  2 \partial_{\bar{z}}\phi^* \partial_z\phi  
-  {\displaystyle\frac{1}{2}}F^* F
+i (\chi ^{\dagger} \partial_z \bar{\chi} + 
\bar{\chi}^{\dagger} \partial_{\bar{z}}\chi).  
\end{equation}

\noindent 
Now, we obtain the matrix $V$ in Eq.(\ref{v}),
\begin{equation}
V = \left(\begin{array}{cl}
  0 & -1 \\
 1 & 0
 \end{array}\right),
\end{equation}

\noindent
from a mass-term in a SUSY-invariant Lagrangian,
\begin{equation}
{\cal{L}}_m  =  - \frac{m}{2} (F^*\phi + F \phi ^* + \chi^{\dagger} \chi - \bar{\chi}^{\dagger} \bar{\chi}) . 
\end{equation}

The na{\"\i}ve lattice SUSY takes of the matrix form
\begin{eqnarray}
Q_L = \left(\begin{array}{cccc}
 0 & 0 & i \epsilon ^* & i \epsilon  \\
 0 & 0 &  - 2 \epsilon   \bigtriangledown_{\bar{z}} & + 2 \epsilon ^*
 \bigtriangledown_z   \\
 - 2 \epsilon ^*  \bigtriangledown_z &  i \epsilon & 0 & 0\\
 - 2 \epsilon   \bigtriangledown_{\bar{z}} & - i \epsilon ^* & 0  & 0 
 \end{array}
 \right), \nonumber 
\end{eqnarray}
\begin{eqnarray}
\end{eqnarray}

\noindent
and 

\begin{eqnarray}
\bar{Q}_L = \left(\begin{array}{cccc}
 0 & 0 & - 2 \epsilon ^* \overleftarrow{\bigtriangledown_z} &  - 2 \epsilon 
 \overleftarrow{\bigtriangledown_{\bar{z}}}  \\
 0 & 0 &  i \epsilon  & - i \epsilon ^* \\
 - i \epsilon ^* & 2
 \epsilon\overleftarrow{\bigtriangledown_{\bar{z}}}
  & 0 & 0  \\
 - i \epsilon &  - 2 \epsilon ^*\overleftarrow{\bigtriangledown_z}
  & 0 & 0 
 \end{array}
 \right). \nonumber
\end{eqnarray}
\begin{eqnarray}
\end{eqnarray}

\noindent
%
It follows that

\begin{eqnarray}
\alpha Q_L + \bar{Q}_L\alpha \nonumber
\end{eqnarray}
\begin{eqnarray}
 = \alpha \left(\begin{array}{cccc}
0 & 0 & -2 \epsilon (TD)_{\bar{z}} & 2 \epsilon^* (TD)_{z} \\
0 & 0 & 0 & 0  \\
 2 \epsilon (TD)_{\bar{z}} & 0 & 0 & 0 \\
-2 \epsilon^* (TD)_{\bar{z}} & 0 & 0 & 0 
 \end{array}
 \right), \nonumber
\end{eqnarray}
\begin{eqnarray}
  \label{td}
\end{eqnarray}

\noindent
where $ (TD)_z$ means  a total derevative
$\overrightarrow{\bigtriangledown_z} + 
\overleftarrow{\bigtriangledown_z}$ .
This result Eq.(\ref{td}) is mentioned in Eq.(\ref{gw2}). 
Therefore, if one uses the symmetric  difference operator,
lattice SUSY  is exactly  same as  na{\"\i}ve one.

Let us derive SUSY extended GW relations.
We denote the lattice effective action $A_{\rm eff}$:

\begin{equation}
A_{\rm eff}[\Psi , \bar{\Psi}] = \sum_{n,m} 
\bar{\Psi}_n S_{(n,m)} \Psi_m,
\end{equation}
\begin{eqnarray}
S = \left(
\begin{array}{ll}
  D_{\rm B} & 0 \\
  0    & D_{\rm F}       
\end{array} \right),
\end{eqnarray}

\noindent
where $D_{\rm B}$ is a bosonic  kinetic part, and
$D_{\rm F}$ is a fermionic one.
From Eq.(\ref{gw2}), we obtain in our approach not only 
the original GW relation but also the relation among fermions 
and bosons:

\begin{equation}
 \sigma_{3}D_{\rm F} + D_{\rm F}\sigma_{3} =
 \alpha^{-1}D_{\rm F}\sigma_{3}D_{\rm F}, \label{sgw1}
\end{equation}
\begin{equation}
 \Sigma_{3}D_{\rm B} + D_{\rm B}\Sigma_{3} =
 \alpha^{-1}D_{\rm B}\Sigma_{3}D_{\rm B},\label{sgw2}
\end{equation}
\begin{eqnarray}
 -2 D^{F^*F}_{\rm B} \overrightarrow{\bigtriangledown}_{\bar{z}} 
 + iD^{\bar{\chi}^{\dagger}\chi}_{\rm F}
 = \alpha ^{-1}D^{F^*F}_{\rm
 B}(TD)_{\bar{z}}D^{\chi^{\dagger}\chi}_{\rm F}, \nonumber
\end{eqnarray}
\begin{eqnarray}
 \label{sgw3}
\end{eqnarray}
\begin{eqnarray}
 i D^{\phi^*\phi}_{\rm B} -2 \overleftarrow{\bigtriangledown}_z 
 D^{\bar{\chi}^{\dagger}\chi}_{\rm F}
 = \alpha^{-1}D^{\phi^*F}_{\rm B}(TD)_z D^{\chi^{\dagger}\chi}_{\rm
 F},
\nonumber \\
 \label{sgw4}
\end{eqnarray}

\noindent
where $\sigma_{3}=\left(\begin{array}{ll}
                           1 & 0 \\ 0 & -1
                        \end{array}\right),
\Sigma_{3}=\left(\begin{array}{ll}
                           1 & 0 \\ 0 & -1
                        \end{array}\right)$,
and the superscript $\chi^{\dagger}\chi$ of 
$D^{\chi^{\dagger}\chi}_{\rm F}$ denotes the $(\chi^{\dagger},\chi)$
component of $D_{\rm F}$.
Eq.(\ref{sgw1}) for fermions is just the original GW relation for 
chiral symmetry, and due to SUSY Eq.(\ref{sgw2}) for bosons was
obtained as the corresponding relation.\cite{ak}
Therefore SUSY can live with $U(1)_{R}$ symmetry on the lattice.
Also Eqs.(\ref{sgw3}),(\ref{sgw4}) represent the supersymmetric
ralation among fermions and bosons.
Once we find the fermion kinetic term
                           $D_{F}^{\bar{\chi}^{\dagger}\chi}$,
it is easy to express the total action explicitly from these relations.

\end{document}